# Two energy gaps observed in tunneling measurements on hole-doped cuprates: pairing gap and coherent gap

A. Mourachkine

*Université Libre de Bruxelles, Service de Physique des Solides, CP233, Boulevard du Triomphe, B-1050 Brussels, Belgium*

There is a clear discrepancy among the energy-gap values for different 90 K cuprates, inferred from tunneling measurements. By using the phase diagram for hole-doped cuprates we show that tunneling measurements performed on 90 K cuprates, simply, detect two different energy gaps: the pairing gap and the coherent gap, which are identical in conventional superconductors. We find that there are two reasons why tunneling measurements show in one cuprate exclusively the coherent gap while, in another cuprate, they show the pairing gap: (i) the number of $CuO_2$ planes per unit cell in the cuprate, and (ii) the directionality of the tunneling current (along *c*-axis or *ab*-plane).

The superconductivity (SC) requires the presence of the Cooper pairs (consisting of either real-space or non-real-space particles) and the long-range phase coherence among the Cooper pairs. In the BCS theory for conventional SCs,[1] the mechanisms responsible for pairing and establishment of the phase coherence are identical: two electrons in each Cooper pair are attracted by phonons, and the phase coherence among the Cooper pairs is established also by phonons. Both phenomena occur almost simultaneously at $T_c$. In SC copper-oxides, there is a consensus that these two mechanisms occur at different temperatures, at least, in the underdoped regime: some kind of pairing exists above $T_c$.[2-6] The magnitudes of energy gaps which correspond to the pairing process (above $T_c$) and establishment of long-range phase coherence (at $T_c$) have different dependencies on hole concentration, $p$, in $CuO_2$ planes.[5,6] The magnitude of the coherent gap, $\Delta_c$, which is proportional to $T_c$, has the parabolic dependence on $p$.[5-7] While the magnitude of the pairing gap, $\Delta_p$, increases linearly with decrease of hole concentration.[5,6,8] Often, the $\Delta_p$ is called a pseudogap which is considered sometimes as a normal-state gap. However, Miyakawa *et al.* showed unambiguously that the pairing $\Delta_p$ gap in $Bi_2Sr_2CaCu_2O_{8+x}$ (Bi2212) is a SC gap.[8] Thus, both the $\Delta_c$ and $\Delta_p$ are SC-like gaps. For instance, in $Tl_2Ba_2CuO_{6+x}$ (Tl2201), there is a clear evidence for the co-existence of two SC components.[9]

It has been widely believed that tunneling measurements are sensitive to probe the $\Delta_p$ rather than the $\Delta_c$.[5] However, one has to note that, for example, tunneling data presented in Ref. 8 show intentionally the maximum magnitudes of tunneling gap in Bi2212 since the d-wave symmetry of the gap has been assumed.[8] In fact, in Bi2212, there is a distribution of the gap magnitude.[10-12] At the same time, the maximum magnitudes of tunneling gaps in Tl2201[13] and $YBa_2Cu_3O_{6+x}$ (YBCO)[14,15] correspond to the minimum gap magnitude in Bi2212.[10-12] However, all three cuprates have similar values of $T_c$. Thus, there is a discrepancy among the energy-gap values for different 90 K cuprates, inferred from tunneling measurements. This problem is often discussed in the literature. In this paper, by using the phase diagram for hole-doped cuprates we show that tunneling measurements performed on Bi2212, YBCO and Tl2201 observe two different energy gaps: the $\Delta_c$ and $\Delta_p$. We find that there are two reason why tunneling measurements show in YBCO and Tl2201 exclusively the coherent $\Delta_c$ gap while, in Bi2212, they show the presence of both the $\Delta_c$ and $\Delta_p$ gaps: (i) the number of $CuO_2$ planes per unit cell in the cuprate, and (ii) the directionality of the tunneling current (along *c*-axis or *ab*-plane).

Figure 1 shows *typical* tunneling spectra[16] obtained on (a) single-layer Tl2201 with $T_c \sim 91$ K;[13] (b) double-layer Bi2212 with $T_c = 89.5$ K,[6] and (c) double-layer YBCO with $T_c = 89$ K (upper curve)[14] and with $T_c = 91$ K (lower curve).[15] The spectra for Tl2201 and YBCO are obtained in SC-insulator-normal metal (SIN) junctions and the two spectra for Bi2212 are obtained in one SC-insulator-SC (SIS) junction. Before analyzing the data it is important to note that (i) these three different cuprates have similar values of $T_c \sim 89 - 91$ K and near optimal doping; (ii) the two spectra shown in Fig. 1(b) are obtained in *one* break junction, *i.e.* on same Bi2212 single crystal, and present *the minimum and maximum gap magnitudes in a junction*, (iii) the data presented in Fig. 2 are obtained on high-quality single crystals, and they are *typical* for more than 100 junctions in each case, and (iv) the tunneling data for Tl2201 and YBCO are obtained along *c*-axis, the data for Bi2212 are measured along *ab*-plane.

In Fig. 1, one can immediately notice the difference between the maximum magnitudes of tunneling gap in Bi2212, on the one hand, and in YBCO and Tl2201, on the other hand. The magnitude of energy gap in Tl2201 and YBCO, $\Delta = 19 - 22$ meV, coincides with the minimum of tunneling gap in Bi2212, $\Delta = 23$ meV. The maximum of tunneling gap in Bi2212 and gap-like features in the density of states (DOS) of the quasiparticle excitations at $V = \pm \Delta_2/e$ in YBCO have similar magnitudes, $\Delta_2 = 30 - 36$ meV.[17]

In order to explain the discrepancy among tunneling data for 90 K cuprates we used the phase diagram for coherent and pairing gaps in hole-doped cuprates,[5,6] shown in Fig. 2. In Fig. 2, the $\Delta_c$ scales with $T_c$ as $2\Delta_c/k_BT_c = 5.45$.[5] The dependence $\Delta_c(p)$ is parabolic since $T_c = T_{c,\,max}[1 - 82.6(p - 0.16)^2]$, where $T_{c,\,max}$ is the maximum $T_c$ for each family of cuprates.[7] It is worth noting that the magnitude of the pairing $\Delta_p$ gap is always larger than the magnitude of the coherent $\Delta_c$ gap.

We found that the values of 19 - 23 meV which correspond to the maximum magnitudes of tunneling gaps in Tl2201 and YBCO and the minimum magnitude of tunneling gap in Bi2212, presented in Fig.1, are in a good agreement with the magnitude of the coherent $\Delta_c$ gap at near optimal doping, shown in Fig. 2. The values of 30 - 36 meV which correspond to the maximum magnitude of tunneling gap in Bi2212 and gap-like features at $V = \pm \Delta_2/e$ in YBCO are in a good agreement with the magnitude of the pairing $\Delta_p$ gap at near optimal doping. So, the reason for the discrepancy among tunneling-gap values for different 90 K cuprates is obvious if we assume that they correspond to two different gaps: $\Delta_c$ and $\Delta_p$. For example, in Bi2212, the presence of two characteristic energy scales have been reported in the literature already a few times. All measurements on near optimally doped Bi2212 show consistent values of the two energy scales, $\Delta_1 = 20 - 23$ meV and $\Delta_2 = 32 - 38$ meV,[10-12,6,8] which are in a good agreement with the $\Delta_c$ and $\Delta_p$ gap values in Fig. 2.

The next step is to find out the reason why tunneling measurements show only the coherent $\Delta_c$ gap in Tl2201 and both the $\Delta_c$ and $\Delta_p$ gaps in Bi2212. The case of YBCO can be considered as an intermediate one between the Tl2201 and Bi2212 cases since tunneling measurements on YBCO show both the $\Delta_c$ and $\Delta_p$ gaps, however, the pairing $\Delta_p$ gap is very weak in tunneling spectra. In Tl2201, as shown by torque measurements, there are also two SC order parameters,[9] however, tunneling measurements show only the coherent $\Delta_c$ gap. It is well known that the tunneling current probes a region of the order of the coherent length $\xi$ from the surface.[18] On the other hand, a torque measurement is the bulk experiment. So, it seems that the reason for the discrepancy among tunneling data performed on 90 K cuprates is the small value of the coherent length in cuprates. Indeed, since $\xi \sim 1/\Delta$,[18] then for the $\Delta_c$ and $\Delta_p$ gaps, the coherent lengths relate to each other always as $\xi_c > \xi_p$ because always $\Delta_c < \Delta_p$ (see Fig. 1). Thus, it is more difficult to probe the pairing $\Delta_p$ gap in surface measurements than the coherent $\Delta_c$ gap since $\xi_c > \xi_p$. In addition to this, in one-layer cuprates, the distance from the surface to the nearest SC $CuO_2$ layer along $c$-axis is larger than in a double-layer compound. Thus, the directionality of the tunneling current is important. For example, tunneling measurements on *one-layer* $Bi_2Sr_2CuO_{8+x}$ performed along CuO$_2$ planes clearly show the presence of the $\Delta_p$ scale.[19] So, we conclude that, in double-layers cuprates, both the $\Delta_c$ and $\Delta_p$ gaps can be observed in tunneling measurements along *c*-axis *and ab*-plane. In one-layer cuprates, tunneling measurements performed along *c*-axis will detect only the coherent $\Delta_c$ gap, and performed along *ab*-plane will show both the $\Delta_c$ and $\Delta_p$ gaps. We predict that tunneling measurements on Tl2201 performed along *ab*-plane, besides the coherent $\Delta_c$ gap, will detect the pairing $\Delta_p$ gap too.

We now turn to the temperature dependencies of the $\Delta_c$ and $\Delta_p$ gaps. In Bi2212, the temperature dependence of tunneling gap having the maximum magnitude (*i.e.* the $\Delta_p$) is almost temperature independent.[2,8] From the common sense, the coherent gap has to diminish to zero at $T_c$ (at least, on the large scale). Figure 3 shows temperature dependencies of two different gaps observed *simultaneously* on same spectrum in a Bi2212 single crystal (see Fig. 3(b) in Ref. 20). Their magnitudes are in a good agreement with the values of the $\Delta_c$ and $\Delta_p$ gaps shown in Fig. 2. Indeed, from Fig. 3, the pairing $\Delta_p$ gap has a tendency to evolve into the pseudogap[2] while the coherent $\Delta_c$ gap decreases with increase of temperature. This is in a good agreement with a MCS (Magnetic Coupling between Stripes) model of the high-$T_c$ SC in hole-doped cuprates, which was presented recently.[12,16,6]

In summary, we discussed the discrepancy among tunneling measurements performed on Tl2201, YBCO and Bi2212. By using the phase diagram for hole-doped cuprates we showed that tunneling measurements performed on 90 K cuprates, most likely, detect two different energy gaps: the pairing $\Delta_p$ gap and the coherent $\Delta_c$ gap. We concluded that there are two reasons why tunneling measurements show in Tl2201 exclusively the coherent $\Delta_c$ gap while, in Bi2212, they show both the $\Delta_c$ and $\Delta_p$ gaps: (i) the number of CuO$_2$ planes per unit cell in the cuprate, and (ii) the directionality of the tunneling current (along *c*-axis or *ab*-plane). The first reason is important due to the small value of the coherent gap in cuprates, which defines the depth of tunneling current.

I thank S. Sergeenkov and R. Deltour for discussions. This work is supported by PAI 4/10.

FIGURE CAPTIONS:

FIG. 1. Tunneling spectra obtained on a single crystal of (a) Tl2201 with $T_c \sim 91$ K,[13] obtained in SIN junction at 4.2 K; (b) Bi2212 with $T_c = 89.5$ K,[6] obtained in one SIS break junction at 14 K, and (c) YBCO with $T_c = 89$ K (upper curve)[14] and $T_c = 91$ K (lower curve),[15] measured in SIN junctions at 10 K and 4.2 K, respectively. The spectra B and C are obtained in *same* junction. The spectrum C does not show the Josephson current since the normal resistance of junction was too high, $R_N = 0.12$ M$\Omega$.[6] The spectra B and E have been shifted vertically for clarity.

FIG. 2. Phase diagram in hole-doped cuprates: $\Delta_c$ is the coherence energy gap, and $\Delta_p$ is the pairing energy gap.[5] The $p_m$ is a hole concentration with the maximum $T_c$.

Fig. 3. Temperature dependencies of quasiparticle DOS measured simultaneously on a Bi2212 single crystal by STM.[20] The solid line corresponds to the BCS temperature dependence. The dashed lines are guides to the eye.

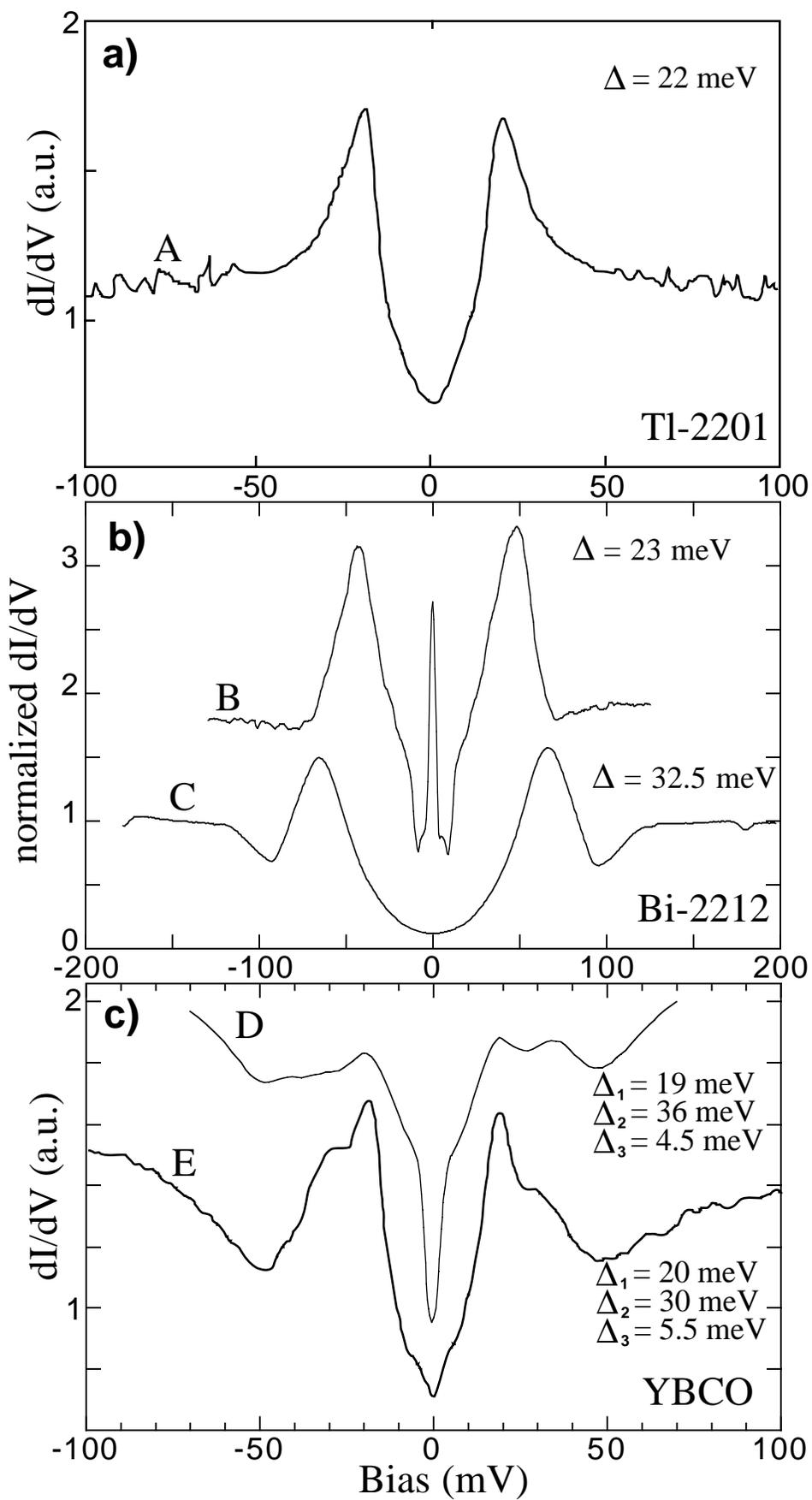

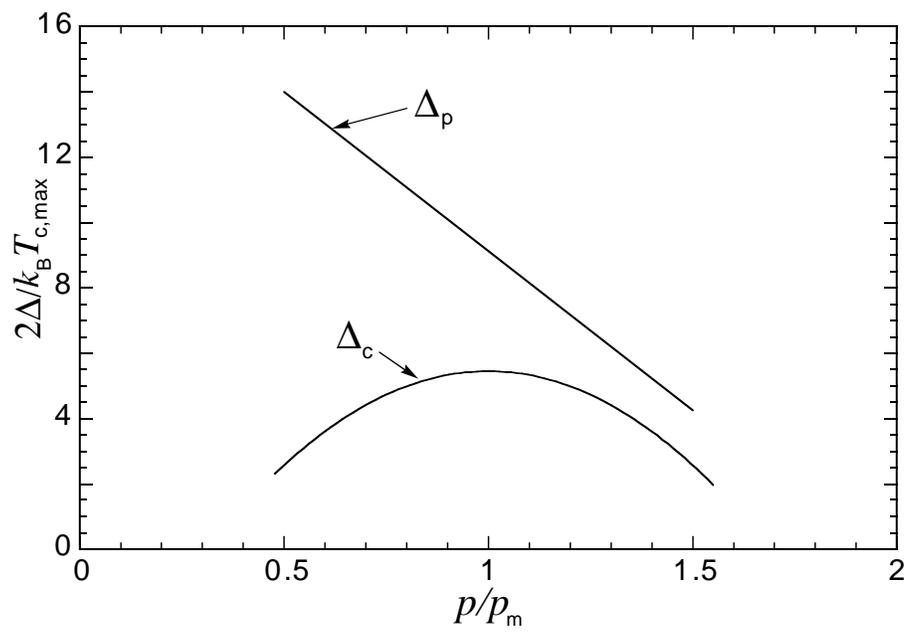

FIG. 2

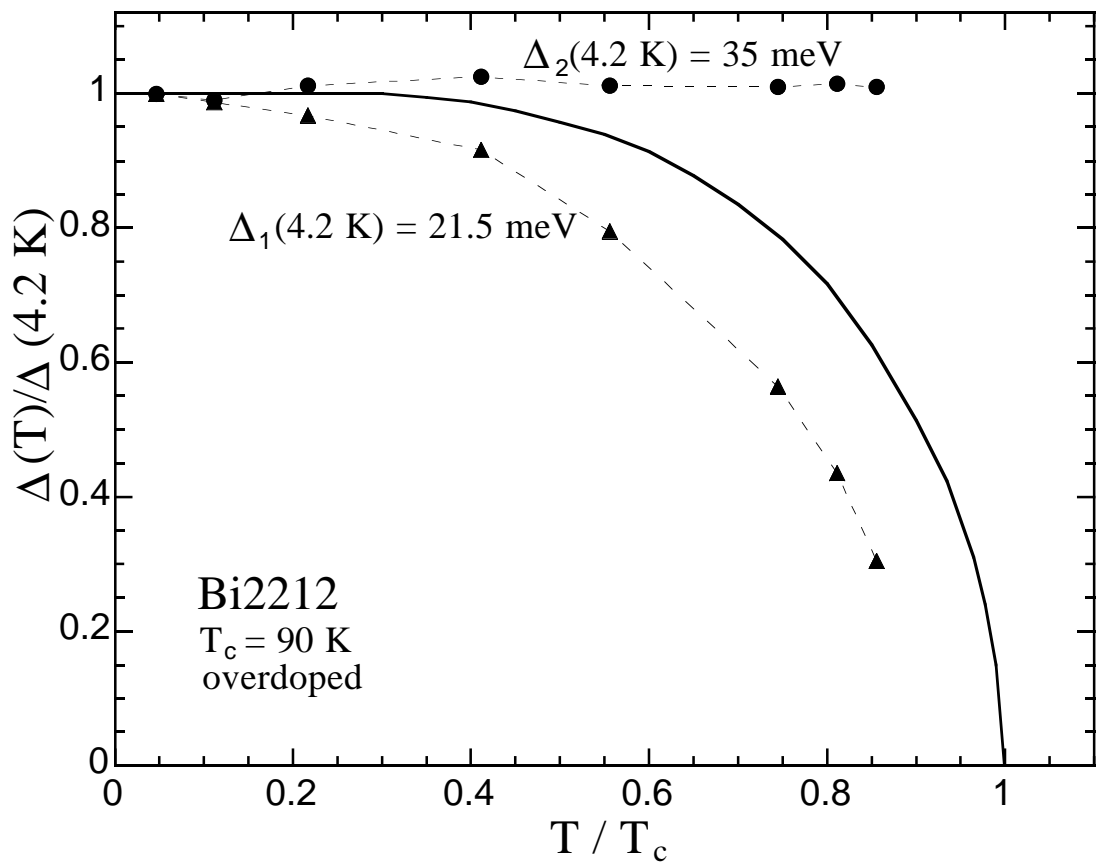

FIG. 3